\documentclass[preprint,showpacs,preprintnumbers,amsmath,amssymb]{revtex4}

\usepackage{slashed}
\usepackage{graphicx}
\usepackage{subfigure}
\usepackage[usenames, dvipsnames]{color}
\usepackage{graphics}
\usepackage{hyperref}
\usepackage{bm}
\usepackage{amsmath}
\usepackage{color}
\usepackage{amsfonts}
\hypersetup{backref,
colorlinks=true,
linkcolor=blue,
linktoc=page,
citecolor=blue,
urlcolor=blue}

\everymath{\displaystyle}

\begin{document}

\title{Quantum Landau damping in dipolar Bose-Einstein condensates}

\author{J. T. Mendon\c{c}a$^1$}

\email{titomend@tecnico.ulisboa.pt}

\author{H. Ter\c cas$^1$}

\email{hugo.tercas@tecnico.ulisboa.pt}

\author{A. Gammal$^2$}

\email{gammal@if.usp.br}

\affiliation{$^1$IPFN, Instituto Superior T\'ecnico, Universidade de Lisboa, 1049-001 Lisboa, Portugal. \\ $^2$Instituto de F\'isica, Universidade de S\~ao Paulo, S\~ao Paulo SP, 05508-090 Brazil.}

\begin{abstract}
We consider Landau damping of elementary excitations in Bose-Einstein condensates (BECs) with dipolar interactions. We discuss quantum and quasi-classical regimes of Landau damping. We use a generalized wave-kinetic description of BECs which, apart from the long range dipolar interactions, also takes into account the quantum fluctuations and the finite energy corrections to short-range interactions. Such a  description is therefore more general than the usual mean field approximation. The present wave-kinetic approach is well suited for the study of kinetic effects in BECs, such as those associated with Landau damping, atom trapping and turbulent diffusion. The inclusion of quantum fluctuations and energy corrections change the dispersion relation and the damping rates, leading to possible experimental signatures of these effects.

Quantum Landau damping is described with generality, and particular examples of dipole condensates in two and three dimensions are studied. The occurrence of roton-maxon configurations, and their relevance to Landau damping is also considered in detail, as well as the changes introduced by the three different processes, associated with dipolar interactions, quantum fluctuations and finite energy range collisions. The present approach is mainly based on a linear perturbative procedure, but the nonlinear regime of Landau damping, which includes atom trapping and atom diffusion, is also briefly discussed. 

\end{abstract}

\maketitle

\section{Introduction}

The study of dipolar systems at low temperature has received considerable attention in recent years \cite{baranov}. In particular, 
the long range interactions between polar atoms and molecules introduces novel aspects of Bose-Einstein condensation, usually 
dominated by contact atomic collisions. In some cases, a convenient manipulation of external fields can nearly remove the 
short-range interatomic forces, and dipolar forces become dominant \cite{koch}. In the usual mean field approximation 
\cite{pethick}, the Gross-Pitaevskii (GP) equation is completed with the inclusion of a non-local interaction potential 
\cite{odell,santos}. But here we use a more general description of the condensates, which is not restricted to mean-field.

Recently, generalized forms of the GP equation have been proposed \cite{fu,sabari}, with cubic and quartic nonlinearities.  The 
usual cubic term describes two-body collisions at zero energy, and the quartic term represents the Lee-Huang-Yang (LHY) correction 
associated with quantum fluctuations \cite{lee,lima}. The LHY correction showed to be essential in explaining the appearance of droplets
in dipolar condensates \cite{pfau}. We also have an additional cubic term resulting from a first order energy 
correction to the two-body collisions \cite{fabrocini}. These new terms can be relevant to the understanding of elementary 
excitations and they damping rates. Rotons have been observed in recent experiments \cite{chomaz}.

Landau damping in BECs has been considered in the past using the mean-field GP equation \cite{stringari,mend2005}, and its relevance to dipole condensates has recently been addressed \cite{natu}. Here we extend the previous analysis to consider both the quantum and quasi-classical regimes, and to discuss the eventual occurrence of atom trapping, quasi-linear diffusion and kinetic instabilities. 
Our approach is also different from previous analysis, because it is not based on the GP equation but makes use of wave-kinetics. This alternative approach is particularly useful for the understanding of kinetic processes, such as those associated with Landau damping.

The WK description of BECs is based on the Wigner function for the quantum medium. This has been explored in the past \cite{gardiner}, and recently extended to condensates at finite temperature \cite{mend2016}.
In this paper, we use a generalized WK equation, which not only includes long range dipolar interactions, but also the effects associated with quantum fluctuations and with a finite energy range of atomic collisions. This can be derived from a generalized GP equation, using a quasi-probability distribution and following the Wigner-Moyal procedure \cite{book}. The inclusion of quantum fluctuations and energy corrections is particularly important to the understanding of the dispersion properties of the elementary excitations in condensates, and to determine the appropriate damping coefficient. This study could therefore lead to possible new experimental signatures of these effects. \par

This paper is organized as follows. In section II, we state the WK equation for dipolar BECs, which is the basic equation of our present model. In section III, we derive the kinetic dispersion relation of elementary excitations in the medium. These excitations are phonon modes of the dipolar quantum fluid. As particular cases, we consider typical configurations, in three (3D) and in quasi two (2D) dimensions. The 3D case contains unstable regions in the range of large wavenumbers, and the quasi-2D case shows the occurrence of a roton-maxon pair \cite{fisher}. In section IV, the kinetic non-dissipative damping of the phonon modes, also known as Landau damping, will be considered. We discuss the cases of a finite temperate BEC, and show that the dipolar interactions modify the Landau damping rate, in both the 3D and quasi-2D configurations. Both quantum and quasi-classical regimes are considered.
We also discuss the possible occurrence of kinetic two-stream instabilities and their relation with the fluid instability studied by \cite{tercas}. We show that Landau damping can still exist for condensates with a finite size, even at zero-temperature. Finite dimensions imply the existence of an residual temperature, as a consequence of the uncertainty principle. This residual temperature is usually very small, but could eventually become relevant near a roton minimum, when the phase velocity approaches zero.  Our discussion of Landau damping is based on of the linearized kinetic equation, and uses the standard perturbative procedure. But, in order to be complete, we discuss in section V, the limits of validity of the linear Landau regime. This discussion includes the main processes that could occur in the nonlinear regime, namely atom trapping and atom diffusion. The atom trapping is a consequence of finite amplitude oscillations, and relies on the possible existence of trapped quantum states. As for atom diffusion, it could occur in the centre-of-mass velocity space due to the presence of a broad spectrum of excitations. Finally, in section VI, we state some conclusions.

\section{Wave-kinetic equation}

 We consider a dipolar condensate, as described by a modified GP equation of the form
\begin{equation}
i \hbar \frac{\partial \psi}{\partial t} = \left( H_{GP} + H' \right) \psi \, , 
\label{2.1} \end{equation}
where $\psi \equiv \psi ({\bf r}; t)$ is the condensate order parameter, describing its ground state, and
$H_{GP}$ is the usual GP Hamiltonian as determined by
\begin{equation}
H_{GP} = - \frac{\hbar^2}{2 m} \nabla^2 + V_0 ({\bf r}) + g \, | \psi ({\bf r}, t)|^2 \, .
\label{2.1b} \end{equation}
Here $V_0 ({\bf r})$ is the confining potential, and $g = 4 \pi \hbar^2 a / m$ is the usual coupling constant describing short-range atomic collisions at zero energy, and $a$ the scattering length. The Hamiltonian $H'$ in eq. (\ref{2.1}) describes three additional effects and can be written as
\begin{equation}
H' = Q |\psi ({\bf r}, t)|^3 + \frac{1}{2} \chi  \left[ \nabla^2 |\psi ({\bf r}, t)|^2 \right] +  \int U_d ({\bf r} - {\bf r'}) | \psi ({\bf r'}, t)|^2 d {\bf r'} \, .
\label{2.1c} \end{equation}
The first term in this expression describes the LHY correction due to quantum fluctuations, determined by the coefficient $Q = g (32/3 \sqrt{\pi}) a^{3/2}$. The second term is due to the finite energy range of atom collisions, and the corresponding coefficient is $\chi=g(a-r_e/2)$, with $r_e$ being the effective range obtained from the second-order expansion of the phase shift \cite{fu}. Finally, the third term describes the long range dipolar interactions and is characterized by a dipole interaction potential $U_d$, to be specified later.

Equation (\ref{2.1}) describes the mean field plus quantum corrections of the condensate wave function $\psi$. In alternative, we can describe the condensate considering the autocorrelation function. This new quantity is usually called the Wigner function, and can be defined as
\begin{equation}
W ({\bf q}, {\bf r}, t) =  \int  \psi^*({\bf r} - {\bf s}/2, t) \psi ({\bf r} + {\bf s}/2, t) \exp(i {\bf q} \cdot {\bf s}) d {\bf s} \, .
\label{2.2} 
\end{equation}
Starting from the above generalized GP equation, and applying the well-known Wigner-Moyal procedure \cite{book}, we can derive an evolution equation for $W$, of the form 
\begin{equation}
i \hbar \left( \frac{\partial}{\partial t} + {\bf v}_q \cdot \nabla \right) W = \int V_k (t) \Delta W \exp(i {\bf k} \cdot {\bf r}) \frac{d {\bf k}}{(2 \pi)^3} \, ,
\label{2.3} 
\end{equation}
where ${\bf v}_q = \hbar {\bf q} / m$ is the atom velocity, and $\Delta W$ is defined as
\begin{equation}
\Delta W = W^- - W^+ \, , \quad W^\pm \equiv W ({\bf q} \pm {\bf k} / 2, {\bf r}, t) \, .
\label{2.3b} 
\end{equation}
The quantity $V_k ( t )$ in Eq. (\ref{2.2}) is the space Fourier transform of the total potential $V ({\bf r}, t)$. We should notice that the integral of the condensate quasi-probability is equal to the local atom density
\begin{equation}
n ({\bf r}, t ) \equiv |\psi ({\bf r}, t)|^2 = \int W ({\bf q}, {\bf r}, t) \frac{d {\bf q}}{(2 \pi)^3}  \, .
\label{2.4} 
\end{equation}
This allows us to write the total potential (\ref{2.1b}) as $V ({\bf r}, t) = [V_0 + g n + Q n^{1/2} + \chi (\nabla^2 n) / 2] + V_d ({\bf r}, t)$, where the dipolar term is determined by
\begin{equation}
V_d ({\bf r}, t) =  \int \frac{d {\bf q}}{(2 \pi)^3} \int d {\bf r'}   U_d ({\bf r} - {\bf r'}) W ({\bf q}, {\bf r'}, t) \, .
\label{2.4b} 
\end{equation}
From the convolution theorem, we have 
\begin{equation}
\int U_d ({\bf r} - {\bf r'}) W ({\bf q}, {\bf r'}, t) \, d {\bf r'} = \int U_d ({\bf k}) W_k ({\bf q}, t) \exp (i {\bf k} \cdot {\bf r}) \,  \frac{d {\bf k}}{(2 \pi)^3} \, ,
\label{2.4c} \end{equation}
where $U_d ({\bf k})$ and $W_k ({\bf q}, t)$ are the space Fourier transforms of the dipolar potential $U_d ({\bf r})$ and the quasi-probability $W ({\bf q}, {\bf r}, t)$. Using this in eq. (\ref{2.4b}), we can transform it into
\begin{equation}
V ({\bf r}, t) =  \int V_k ( t ) \exp (i {\bf k} \cdot {\bf r}) \,  \frac{d {\bf k}}{(2 \pi)^3}    \, ,
\label{2.5} 
\end{equation}
with
\begin{equation}
V_k ( t ) = V_0 ({\bf k}) + \left[ g - \frac{k^2}{2} \chi + U_d ({\bf k}) \right]Ên_k ( t ) + Q \int n^{3/2} ({\bf r}, t)  \exp (- i {\bf k} \cdot {\bf r}) d {\bf r} \, ,
\label{2.5b} \end{equation}
and $n_k ( t)$ is the spectral component of the BEC density, as given by the space Fourier transform of Eq. (\ref{2.4}). The wave-kinetic equation in (\ref{2.3}), together with the expression for $V_k ( t )$ in Eq. \eqref{2.5b}, provides the full phase-space descprition of a dipolar BEC in the presence of quantum corrections.

\section{Dispersion relation}

In order to discuss the elementary excitations of the dipolar BEC, we assume that the quasi-probability can be divided in two distinct parts,  $W = W_0 + \tilde W$. Here, $W_0$ is the equilibrium distribution describing the condensate in steady state, and $\tilde W$ is a small perturbation such that $| \tilde W | \ll | W_0 |$, describing the elementary excitations of the medium. Let us consider the simple case of a uniform and unbounded medium and assume a plane wave  perturbation of the form
\begin{equation}
\tilde W ({\bf q}, {\bf r}, t) = \tilde W_k ({\bf q}) \exp (i {\bf k} \cdot {\bf r} - i \omega t) \, ,
\label{3.1} 
\end{equation}
where $\omega$ is the mode frequency. The corresponding density perturbation will be
$\tilde n ({\bf r}, t )  =  \tilde n_k  \exp (i {\bf k} \cdot {\bf r} - i \omega t) $.
Linearizing Eq. (\ref{2.3}) with respect to the perturbed quantities, we can then easily get
\begin{equation}
\tilde W_k = \left[g + Q \sqrt{n_0} - \frac{\chi}{2} k^2  + U_d ({\bf k}) \right] \frac{\Delta W_0}{\hbar (\omega - {\bf k} \cdot {\bf v}_q)}\,Ê\tilde n_k  \, .
\label{3.2} 
\end{equation}
Integrating over the atom momentum, we can then obtain a dispersion relation of the form
\begin{equation}
1 -  \left[g + Q \sqrt{n_0} - \frac{\chi}{2} k^2  + U_d ({\bf k}) \right]  \int \frac{\Delta W_0}{\hbar (\omega - {\bf k} \cdot {\bf v}_q)}\,Ê\frac{d {\bf q}}{(2 \pi)^3}  = 0  \, .
\label{3.2} 
\end{equation}
This is valid for any condensate with long-range dipolar interactions. The latter can also be written as
\begin{equation}
1 -  \left[g + Q \sqrt{n_0} - \frac{\chi}{2} k^2  + U_d ({\bf k}) \right] \int \frac{W_0 ({\bf q})}{\hbar} \left[ \frac{1}{(\omega_- - {\bf k} \cdot {\bf v}_q)} - \frac{1}{(\omega_+ - {\bf k} \cdot {\bf v}_q)} \right] Ê\frac{d {\bf q}}{(2 \pi)^3}  = 0  \, ,
\label{3.2b} 
\end{equation}
with $\omega_\pm = \omega \pm \frac{\hbar k^2}{2 m}$. We first study the dispersion relation for a zero-temperature BEC. This allows us to use the simple equilibrium distribution $W_0 ({\bf q}) = (2 \pi)^3 n_0 \delta ({\bf q} -{\bf q}_0)$, where $n_0$ is the unperturbed density and ${\bf q}_0$ defines a constant drift velocity ${\bf v}_0 = \hbar {\bf q}_0 / m$. Eq. (\ref{3.2b}) is then reduced to 
\begin{equation}
1 -  \left[g + Q \sqrt{n_0} - \frac{\chi}{2} k^2  + U_d ({\bf k}) \right] \frac{n_0}{\hbar} \left[ \frac{1}{(\omega_- - {\bf k} \cdot {\bf v}_0)} - \frac{1}{(\omega_+ - {\bf k} \cdot {\bf v}_0)} \right] Ê{(2 \pi)^3}  = 0  \, .
\label{3.3} \end{equation}
Rearranging therms and using the Bogoliubov speed, $c_s = \sqrt{g n_0 / m}$, this can also be written as
\begin{equation}
\left( \omega - {\bf k} \cdot {\bf v}_0 \right)^2 = k^2 c_s^2 \left[1 + \frac{Q}{g} \sqrt{n_0} -  \frac{\chi}{2 g} k^2  + \frac{1}{g} U_d ({\bf k}) \right] + \frac{\hbar^2 k^4}{4 m^2} \, .
\label{3.3b} \end{equation}
This can easily be extended to the case of two counter-propagating BEC beams \cite{tercas}. 
It is useful to consider the product of the phase velocity $v_\phi = \omega / k$ and the group velocity $v_g = \partial \omega / \partial k$. Assuming a condensate at rest ($v_0 = 0$), we obtain
\begin{equation}
v_\phi v_g = c_s^2 \left[1 + \frac{Q}{g} \sqrt{n_0} + \frac{1}{g} U_d ({\bf k}) \right] + 4 k^2 \left( \frac{\hbar^2}{4 m^2} - \frac{\chi}{2 g} \right).
\label{3.3c} 
\end{equation}
This shows that, the product $v_\phi v_g$ is nearly equal to the square of the Bogoliubov speed, with corrections coming from the quantum dispersion term and from the three different processes included in the present model (dipolar potential, quantum fluctuations and finite energy range of close collisions).
We can now take particular examples of dipolar potential. For typical dipole condensates, we can use the long range interaction potential  \cite{giovanazzi,kumar}
\begin{equation}
U_d ({\bf R}) = \frac{C_{dd}}{4 \pi} \frac{1 - 3 \cos^2 \theta}{| {\bf R}|^3} \left(\frac{3 \cos^2 \varphi - 1}{2} \right) \, ,
\label{3.4} \end{equation} 
where $C_{dd}$ is the magnetic dipole constant, $\theta$ is the angle between the vector difference ${\bf R} = {\bf r} - {\bf r'}$ and the direction of the external polarization field, and $\varphi$ is the angle between the orientation of the dipoles and the $z$-axis. Taking $\varphi = 0$, we obtain the Fourier transform
\begin{equation}
U_d ({\bf k}) = C_{dd} \left( \cos^2 \theta_k - \frac{1}{3} \right) \, ,
\label{3.4b} \end{equation}
where $\theta_k$ is the angle between the wavevector ${\bf k}$ and the $z$-axis. Replacing this in eq. (\ref{3.3b}), and assuming a condensate at test (${\bf v}_0 = 0$), we obtain the dispersion relation
\begin{equation}
\omega^2 = k^2 c^2 (\theta_k) + \frac{\hbar^2 k^4}{4 m^2} \, .
\label{3.5} \end{equation}
where  $c (\theta_k)$ is the angle dependent Bogoliubov velocity, defined by
\begin{equation}
c^2 (\theta_k) = c_s^2 \left[ 1 + \frac{Q}{g} \sqrt{n_0} -  \frac{k^2}{2 g} \chi  + \eta \left( \cos^2 \theta_k - \frac{1}{3} \right) \right] \, .
\label{3.5b} \end{equation}
Here, $\eta = C_{dd} / g$ is the ratio between the dipole and contact potential strength.
As we can see, the dipole interactions introduce important qualitative corrections to the characteristic sound velocity, which can become imaginary for large values of the parameter $\eta$.  In particular, a critical wavenumber $k_c$ can be defined, where $\omega^2 = 0$, as  
\begin{equation}
k_c^2 = c_s^2 \left[ \eta \left( \frac{1}{3} - \cos^2 \theta_k \right) - \left( 1 + \frac{Q}{g} \sqrt{n_0} \right) \right]  
\left(  \frac{4 m^2}{\hbar^2} - \frac{\chi}{2 g} \right) \, .
\label{3.6} 
\end{equation}
This is positive for $\theta_k \simeq \pi/2$ and $\eta \geq  3$. In such case, large scale perturbations corresponding to $k \leq k_c$ become unstable, with a finite growth rate determined by $\omega^2 \leq 0$. This is physically relevant for $( k_c a) \geq 1$, where $a$ is the typical size of the condensate.

Another interesting example is the quasi-2D condensate. If a BEC is strongly confined along the $z$-axis, which size $l_z$ much small than its  transverse dimension $a$, we can still use the same WK equation, only depending on $(x, y)$ and $(k_x, k_y)$, but  where $g$ is replaced by a renormalized coupling parameter, $g_{2D} = g / (3\sqrt{2 \pi} l_z)$. In this case, the quasi-2D dipole interaction potential can be represented by \cite{natu}
\begin{equation}
U_d (k) = C_{dd} F (k l_z / \sqrt{2})  \, ,
\label{3.6} \end{equation}
where $k = \sqrt{k_x^2 + k_y^2}$ and the function $F (x)$, with $x = k l_z / \sqrt{2}$,  is defined as
\begin{equation}
F (x)  = 1 - \frac{3}{2} \sqrt{\pi} x \exp(x^2) \, \mathrm{erfc} ( x) \, .
\label{3.6b} \end{equation}
Here, we have used the complementary error function, defined by
\begin{equation}
\mathrm{erfc} ( x ) = \frac{2}{\sqrt{\pi}} \int_x^\infty \exp(- t^2) \, dt \, .
\label{3.6c} \end{equation}
For our discussion, it is useful to consider the asymptotic expansion  for $x \ll 1$, or $k \ll \sqrt{2} / l_z$, as 
$\mathrm{erfc} ( x ) \simeq 1 - x ( 1 - x^3 / 3) / \sqrt{\pi}$. 
Using this new dipole potential in eq. (\ref{3.3b}), we can then write
\begin{equation}
\omega^2 = k^2 c_{2D}^2 \left[ 1 + \frac{Q}{g} \sqrt{n_0} -  \frac{k^2}{2 g} \chi + \epsilon_{dd} F (k l_z / \sqrt{2}) \right] + \frac{\hbar^2 k^4}{4 m^2}  \, ,
\label{3.7} \end{equation}
where we have defined the new quantities
$c_{2D}^2 = g_{2D} ( n_0 / m )$, and  $ \epsilon_{dd} = C_{dd} / g_{2D}$.
For $k l_z \ll \sqrt{2}$, we can use the approximate expression
\begin{equation}
\omega^2 = k^2 c_{2D}^2 \left\{ 1 + \frac{Q}{g} \sqrt{n_0} -  \frac{k^2}{2 g} \chi + \epsilon_{dd} \left[ 1 - \frac{k l_z}{\sqrt{2 \pi}} \left(1 -  \frac{k^3 l_z^3}{3 \cdot 2^{3/2}}  \right) \right] \right\} + \frac{\hbar^2 k^4}{4 m^2}  \, .
\label{3.8} \end{equation}
It is well known that these dispersion relations can lead to the occurrence of a roton-maxon pair. In some extreme conditions, this can even lead to the formation of a super-solid, where $\omega^2$ becomes negative for a well defined wavenumber $k$ (and not over a large region $0 \leq k \leq k_c$, as in the above 3D example). A necessary super-solid condition is $\omega^2 = 0$, at a critical value $k = k_0$, such that
\begin{equation}
F (k_0 l_z / \sqrt{2}) \simeq -  \frac{1}{\epsilon_{dd}} \left( 1 + \frac{\hbar^2 k_0^2}{4 m^2} \frac{1}{c_{2D}^2} \right) \,.
\label{3.9} \end{equation}
In this expression, we have neglected the terms in $Q$ and $\chi$, which that have been recast in Fig. \ref{fig1} for full illustration. Comparing this with Eq. (\ref{3.7}), we can see that the super-solid instability cannot occur for small values of $k \ll 1/l_z \sqrt{2}$, but will eventually exist in the region of large wavenumbers.

\begin{figure}
\includegraphics[angle=0,scale=0.8]{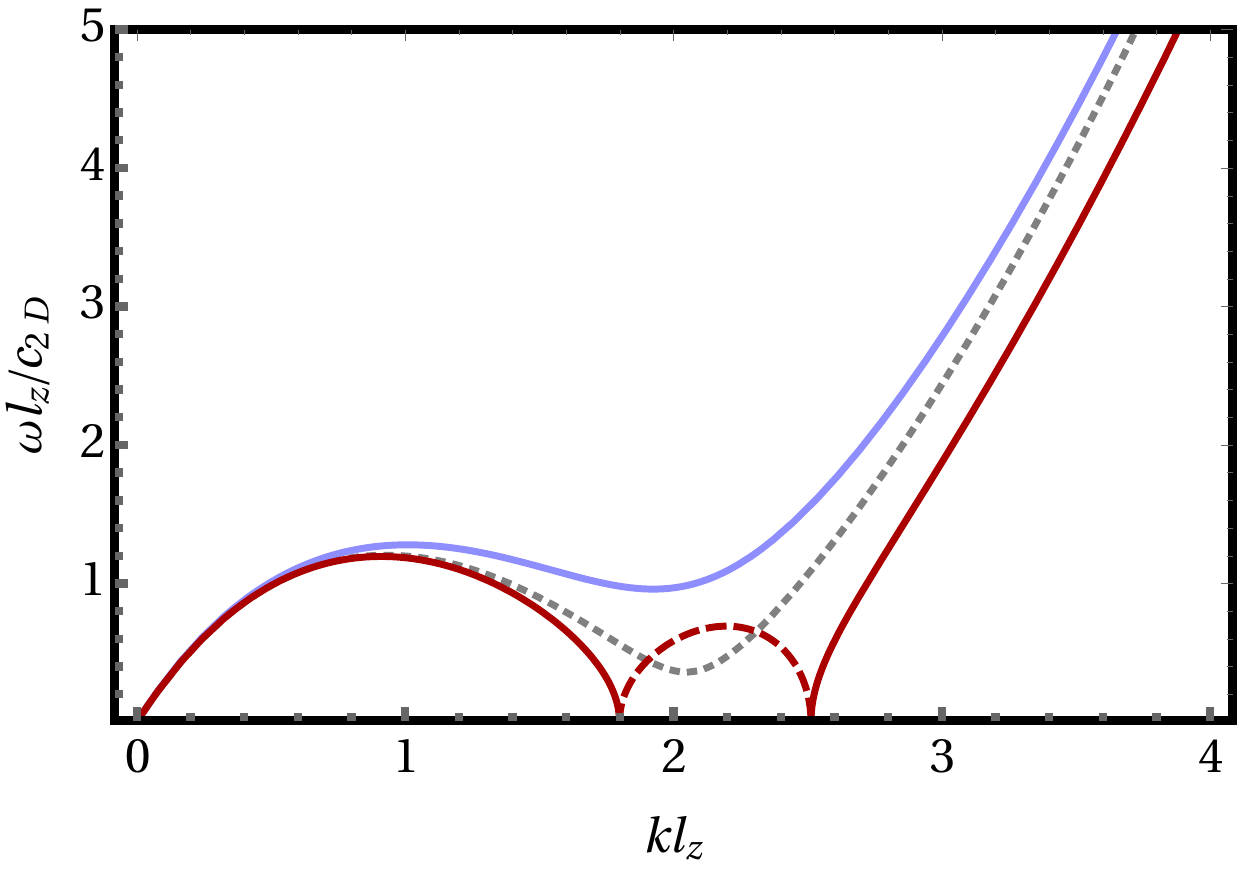}
\caption{ (color online) Dispersion relation of elementary phonon modes in a quasi-2D dipolar condensate, displaying a roton-maxon configuration. The phase velocity $v_\phi = \omega / k$ decreases in the roton region and increases in the maxon region, thus changing the value of the corresponding Landau damping rates. The dotted line depicts the mean-field situation ($Q=0$ and $\chi=0$). The stabilization of the roton instability due to the LHY quantum correction is represented for $Q=0.2g/\sqrt{n_0}$ (blue solid line). The inclusion of the finite range of the atom collisions enhance the roton stability, here depicted for $\chi=0.08 g/l_z^2$ (solid red line). The dashed line represents the imaginary part of $\omega$ in Eq. \eqref{3.8}. In all situations, we have set $\epsilon_{dd}=8.2$. }
\label{fig1}
\end{figure}

\section{Landau damping}

The wave-kinetic description is particularly well suited to describe Landau damping and the related kinetic instabilities, as shown next. For that purpose, we go back to eq. (\ref{3.2b}), which can be rewritten in the form
\begin{equation}
1 -  \frac{g' (k)}{\hbar k}  \int G_0 (u) \left[ \frac{1}{(u - \omega_+ / k)} - \frac{1}{(u - \omega_- / q)} \right] Ê\frac{d u}{2 \pi}  = 0  \, ,
\label{4.1} \end{equation}
with $g' (k ) = [g + Q \sqrt{n_0} - k^2 \chi / 2 + U_d ( k )]$. Here, $u$ and $q$ represent the atom velocity and momentum components parallel to the direction of propagation,according to. 
\begin{equation}
{\bf v}_q = u \frac{{\bf k}}{k} + {\bf v}_\perp \, , \quad {\bf q} = q \frac{{\bf k}}{k} + {\bf q}_\perp \, .
\label{4.1b} \end{equation}
We have also used the reduced distribution $G_0 ( q)$, such that 
\begin{equation}
G_0 ( q) = \int W_0 (q, {\bf q}_\perp) \frac{d {\bf q}_\perp}{(2 \pi)^2} \, .
\label{4.2} \end{equation}
The integrals in eq.  (\ref{4.1}) includes two integrals, can be written in the form
\begin{equation} 
\int  \frac{G_0 (u) }{(u - v_\pm)}  du = \mathcal{P} \int  \frac{G_0 (u) }{(u - v_\pm)}  du + i \pi G_0 (v_\pm) \, .
\label{4.2b} \end{equation}
where $v_\pm = \omega_\pm / k$,  and $\mathcal{P} $ represents the principal part of the integral, in the Cauchy sense. Using this in eq. (\ref{4.1}) we can write it in the form $\epsilon (\omega, k)  = 0$, which can be split into its real and imaginary parts, as $\epsilon = \epsilon_r + i \epsilon_i$. For a real value of $k$, this leads to a complex mode frequency $\omega = \omega_r + i \gamma$. Assuming 
$| \gamma | \ll \omega_r$, we can determine separately the frequency $\omega_r$ and the damping rate $\gamma$, by writing
\begin{equation}
\epsilon_r (\omega_r, k) = 0 \, , \quad \gamma = - \frac{\epsilon_i (\omega_r, k)}{(\partial \epsilon_r / \partial \omega)_{\omega_r}} \, .
\label{4.3} \end{equation}
Temperature effects are usually be negligible in what concerns the value of $\omega_r$. We are then allowed to use $G_0 (u) = 2 \pi G_0 \delta ( u )$ in the first of these equations, which then reduces to
\begin{equation}
\epsilon_r (\omega_r, k) = 1 - \frac{g' ( k) k^2}{m} \frac{G_0}{\omega^2 - \hbar^2 k^4 / 4 m^2} = 0 \, .
\label{4.4} \end{equation}
Writing $G_0 = n_0$ this  reduces to eq. (\ref{3.3b}), for a BEC at rest. We can then approximately write
\begin{equation}
\frac{\partial \, \epsilon_r}{\partial \omega} = \frac{2 g \omega}{k^2 c_s^2 g' (k)} \, .
\label{4.4b} \end{equation}
Retaining finite temperature effects in the damping rate (\ref{4.3}), we can then obtain
\begin{equation}
\gamma = \frac{g k c_s^2}{4 \hbar \omega_r} \left[ 1 + \frac{Q}{g} \sqrt{n_0} -  \frac{k^2}{2 g} \chi  + \frac{1}{g} U_d (k ) \right]^2 \left[G_0 (v_+) - G_0 (v_-) \right] \, .
\label{4.5} \end{equation}
This determines the atomic Landau damping of elementary excitations in dipolar condensates. In thermal equilibrium, we always have $G_0 (v_+) < G_0 (v_-)$, and the damping coefficient is negative, $\gamma < 0$. But, in a disturbed BEC, an inversion of population can eventually occur, such that $G_0 (v_+) > G_0 (v_-)$. In this case the excitations are kinetically unstable. It is important to notice that the quantum fluctuations associated with $Q$, and the finite energy collisions described by $\chi$, never change the sign of $\gamma$ in the above expression, because  the damping rate only depends on the quantity $g' (k)^2$. \par

It is also useful to consider the semiclassical limit, valid for $| k | \ll | q |$. In this case, we can develop the quantities $G_0 (v_\pm)$ around $v = \omega / k$, and eq. (\ref{4.5}) becomes
\begin{equation}
\gamma \simeq \frac{k^3 c_s^4}{4 n_0 \omega_r} \left[ 1 + \frac{Q}{g} \sqrt{n_0} -  \frac{k^2}{2 g} \chi  + \frac{1}{g} U_d (k ) \right]^2 \left( \frac{\partial G_0}{\partial v} \right)_{v = \omega / k} \, .
\label{4.6b} \end{equation}
For a condensate in equilibrium at a finite temperature $T$, the derivative is always negative and the excitations are damped. In order to be more specific, we need an explicit expression for the reduced distribution $G_0 (v)$. We can use a Bose-Einstein distribution.
\begin{equation}
G_0 (v) = 2 \pi n_0 \left\{ e^{[E(v) - \mu] \beta} - 1 \right\}^{-1}
\label{4.7} \end{equation} 
where $E (v) = m v^2 / 2 = \hbar^2 q^2/ 2 m$, $\beta = 1 / k_B T$, and the chemical potential $\mu$ provides the zero of the energy scale. The Landau damping rate $\gamma$ becomes
\begin{equation}
\gamma \simeq - \frac{k^3 c_s^4}{4 \omega_r} \left[ 1 + \frac{Q}{g} \sqrt{n_0} -  \frac{k^2}{2 g} \chi  + \frac{1}{g} U_d (k ) \right]^2 \beta e^{[E(\omega/k) - \mu] \beta}\left\{ e^{[E(\omega/k) - \mu] \beta} - 1 \right\}^{-2}  \, .
\label{4.7b} \end{equation}
We can see that $\gamma$ is always negative, for all possible values of the phase velocity $\omega / k$. However, out of equilibrium situations can eventually occur, where the BEC is kinetically unstable. This is linked with the possible existence of a supra-thermal atomic stream, with density $n_b$ and mean velocity $v_b$, as described by 
\begin{equation}
G_0 (v) = \frac{2 \pi  n_0}{e^{[E(v) - \mu] \beta} - 1} +  \frac{2 \pi n_b}{e^{[E(v - v_b)) - \mu] \beta_b} - 1} \, ,
\label{4.8} \end{equation} 
where $\beta_b = 1 / k_B T_b$ and $T_b$ is the temperature of the supra-thermal stream. In this case, the sign of the mode damping coefficient $\gamma$ will eventually change sign, leading to an unstable region of phase velocities $| v | < \omega / k \leq | v_b|$. This is the kinetic counterpart of the two-stream instability discussed in \cite{tercas}. 
 
\begin{figure}
\includegraphics[angle=0,scale=0.75]{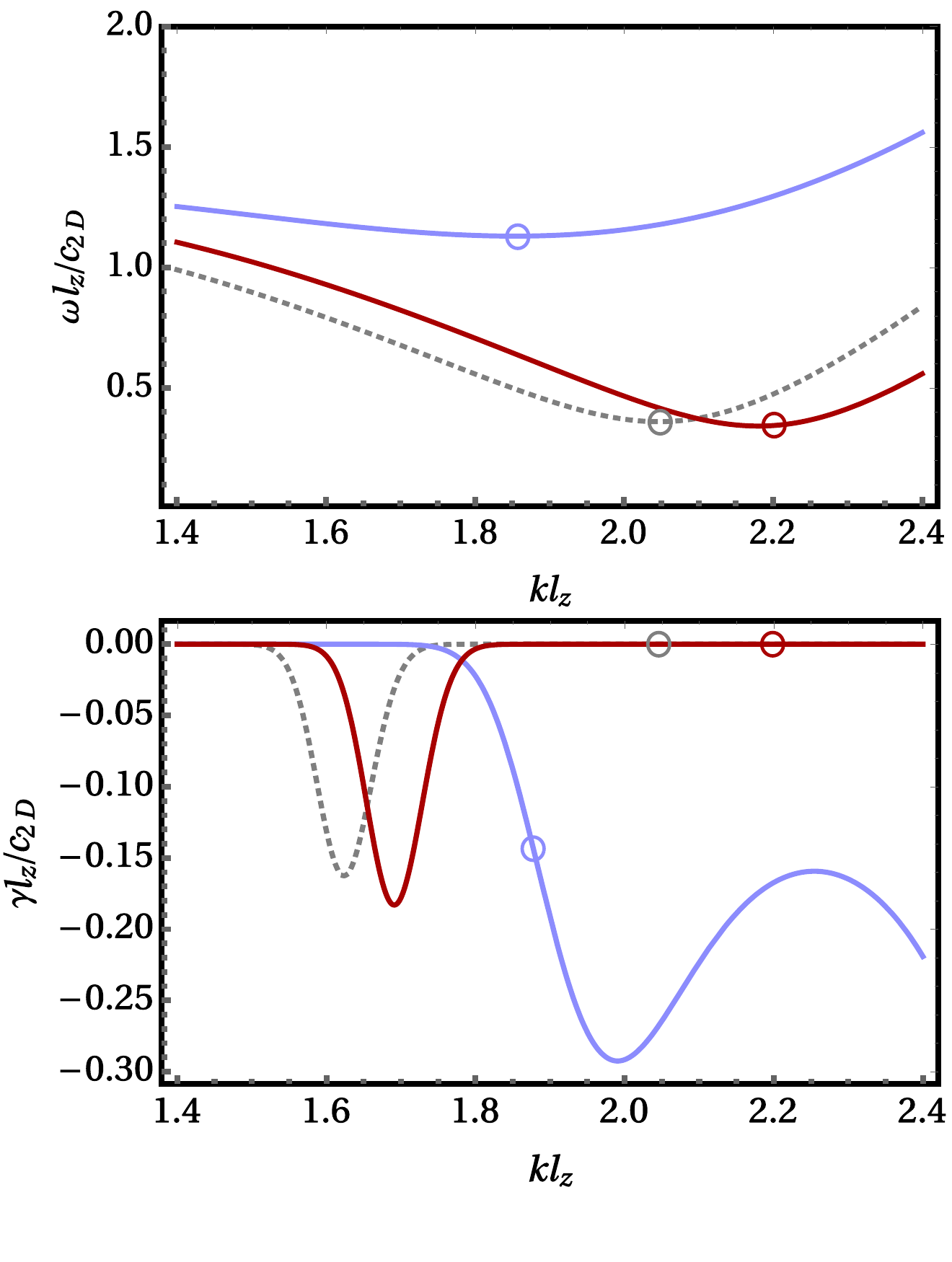}
\caption{(color online) Estimation of the quantum Landau damping due to the Heinsenberg uncertainty. Real (top) and imaginary (bottom) parts of the dispersion relation in Eq. \eqref{3.7}. Depicted are the mean-field case $Q=\chi=0$ (dashed line), the LHY correction, $Q=0.3g/\sqrt{n_0}$ and $\chi=0$ (light solid line) and its combination with the finite range of the atomic collisions, $Q=0.3g/\sqrt{n_0}$ and $\chi=0.08 g/l_z^2$ (dark solid line). The circles mark the position of the roton minimum in the different situations. In all cases, we have set $\epsilon_{dd}=8.2$ and a small residual temperature of $T_r=0.01 m c_{2D}^2/k_B$, which is much less than the critical temperature for condensation $T_c$.}
\label{fig_damping}
\end{figure}

Finally, it is important to notice that, even a zero temperature $T = 0$, the Landau damping may happen. This is due to the uncertainty principle, which implies that for a BEC with typical size $L$ the uncertainty of the atom velocity will be $\Delta v \simeq \hbar / m L$. Therefore, a finite size $L$ is equivalent to a residual temperature $T_r$ of order of $\Delta v^2$. In that case, the Heisenberg broadens the zero-temperature distribution $G_0(v)$ with the residual temperature 
\begin{equation}
T \rightarrow T_r = \frac{1}{2 k_B} \frac{\hbar^2}{m L^2}
\label{4.8b} 
\end{equation}
For a Dy BEC, with $m\sim 162$ au, chemical potential $\mu\sim 3$ kHz and $L\sim 10-100$ $\mu$m, we obtain $T_r\sim 0.3-30$ pK, much less than the critical temperature $T_c\sim 80$ nK \citep{dysprosium1, dysprosium2}. This means that Landau damping will mainly be provided by the thermal part of  the condensed gas. However, in a situation where the phase velocity $\omega / k$ of the elementary excitation is strongly reduced in the viscinity of a roton minimum, Landau damping could eventually be provided by the condensed gas itself, due to the existence of a residual temperature $T_r$. This feature is illustrated in Fig \ref{fig_damping}. As we can see, in the mean-field case, Landau damping occurs below the roton minimum, while the inclusion of the quantum LHY correction displaces the Landau damping towards the roton minimum. With the inclusion of the finite range of the atomic collisions, the roton minimum remains practically undamped.

\begin{figure}
\includegraphics[angle=0,scale=0.75]{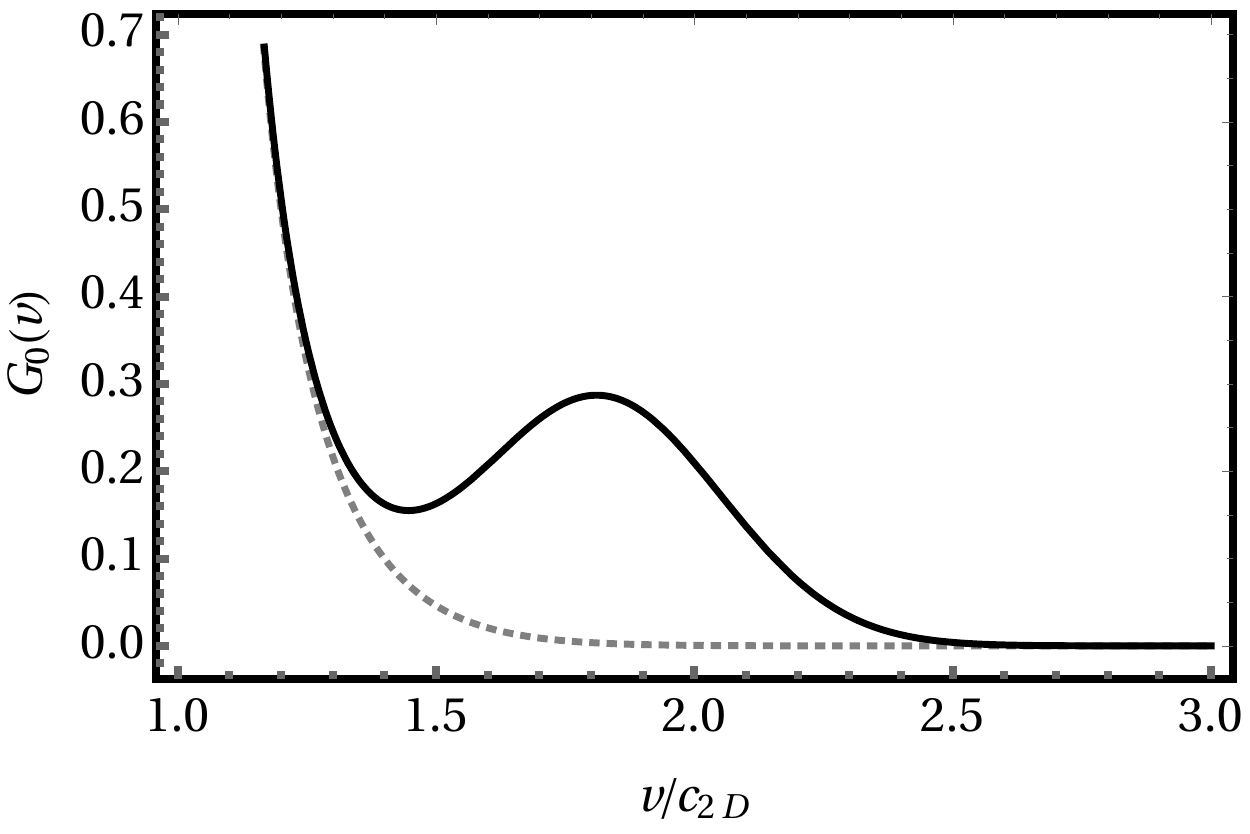}
\caption{\label{fig1}  Unstable reduced distribution $G_0 (v)$, of a thermal dipolar condensate in the presence of a stream of atoms (solid line). For comparison, a stable thermal distribution is depicted (dashed line).}
\end{figure}

\section{Trapping and diffusion}

Landau damping results from an energy exchange between the phonon excitations and the atomic mean field. The above linear description of Landau damping needs to be completed by a discussion of other possible effects associated with this energy exchange. The first is atomic trapping, which can take place for an oscillation with a finite amplitude. Another is atom diffusion, when a large spectrum of excitations is excited in the medium. In this case, the exchange of energy between the mean field and the phonon field induces diffusion in the atomic velocity space, associated with the cumulative Landau damping over the photon spectrum, with second-order changes in the mean field. These two aspects, trapping and diffusion, are briefly described next. In this section, we will neglect quantum fluctuations and finite energy range effects, and take $Q = 0$ and $\chi = 0$, for simplicity. In the discussion that follows, the latter effects only as small corrections. 

Atom trapping occurs in the close vicinity of resonance, when the atom velocity $v$ is equal to the phonon phase velocity. It can easily be seen that the centre-of-mass energy of the trapped  states falls in the range $E_{res} - g \tilde n_k \leq E_{trap} \leq E_{res} + g \tilde n_k$, where $\tilde n_k$ is the phonon amplitude and $E_{res} = \hbar^2 \omega^2/ 2 m k^2$. This means that the trapped states correspond to wavenumbers $q$ in the interval $q_- \leq q \leq q_+$, where
\begin{equation}
q_\pm \simeq \frac{\omega}{k} \sqrt{1 \pm \frac{k^2 c_s^2}{\omega^2} \frac{2 m^2}{\hbar^2} \left( \frac{\tilde n_k}{n_0} \right)}
\label{5.1} \end{equation}
The potential well created by the phonon excitations creates a series of trapped energy states, with energies levels  $\hbar \omega_B (1/2 + \nu)$, not exceeding  $g \tilde n_k$, where $\nu$ is an integer. The {\sl bounce frequency} for the trapped atoms is given by
\begin{equation}
\omega_B = k c_s \sqrt{\frac{\tilde n_k}{2 n_0} \left[ 1 + \frac{1}{g} U_d (k) \right] } \, .
\label{5.2} \end{equation}
This trapping process is very similar to that occurring for free electrons in a  quantum plasmas \cite{brodin,daligault}, even if the atoms in a BEC are bosons and the electrons in a plasma are fermions. In particular, we can define a similar {\sl trapping parameter}, $R_{trap} = g \tilde n / \hbar \omega_B$, which gives the approximate number of trapped states. For $R_{trap} \gg 1$, we are in the quasi-classical limit, and for $R_{trap} < 1/2$ trapping will be forbidden. Trapping introduces nonlinear corrections to Landau damping, which can lead to modulations of the mode amplitude at the harmonics of the bounce frequency $\omega_B$. However, nonlinear Landau damping is outside the scope of the present work.

Let is now consider the case of a broad spectrum of phonons. This is relevant to a turbulent BEC. A quasi-linear theory, based on the above wave-kinetic equation can then be establish, which is formally identical to that derived in \cite{mend2010} for a laser-cooled gas. Each phonon excitation will be damped with the corresponding Landau damping rate, but due to global energy transfer between the mean field and the turbulent field, the equilibrium distribution $W_0 ({\bf q})$ will change over a long time-scale, as determined by the diffusion equation
\begin{equation}
\left[ \frac{\partial}{\partial t} + {\bf v}_q \cdot \nabla - \frac{\partial}{\partial {\bf q}} \cdot \bar{\bar {\bf D}} \cdot \frac{\partial}{\partial {\bf q}} \right] W_0 ({\bf q}, t) = 0 \, ,
\label{5.3} \end{equation}
were $\bar{\bar {\bf D}}$ is a diffusion tensor in the atomic velocity space, given by
\begin{equation}
{\bar{\bar {\bf D}}} = \frac{\pi}{n_0^2}  \int \left[1  + \frac{1}{g} U_d ( {\bf k}) \right]^2 {\bf q} {\bf q}  \, | \tilde n_k |^2 \, \delta (\omega - {\bf k} \cdot {\bf v}_q ) \frac{d {\bf k}}{(2 \pi)^3}
\label{5.3b} \end{equation}
This expression shows that diffusion results from the accumulation of resonant interactions of the centre-of-mass states with the different Fourier components of the phonon spectrum. A detailed study of the diffusion equations is outside the scope of the present work.
 
\section{Conclusions}

In this paper, we have described main properties of quantum Landau damping in dipolar condensates. The quasi-classical limit was also discussed. Our model was based on a generalized wave-kinetic equation, with a non-local potential, where quantum fluctuations and the finite energy corrections were also included . We have shown that such a kinetic description is particularly adequate to describe Landau damping and kinetic instabilities associated with deviations from thermal equilibrium. 

A general expression for the dispersion relation of elementary excitations in the dipolar BEC, and the corresponding Landau damping rate, were established. Typical dipolar configurations in three and quasi-two dimensions were also examined, which included the formation of maxon-roton pairs and the eventual occurrence of supersolids. Landau damping tends to increase in the presence of a maxon-roton pair, because the roton minimum decrease the phonon phase velocity, bringing the resonant phonon-atom interaction closer to the thermal velocity. The opposite situations occurs for near the maxon region. 

Possible kinetic instability regimes were discussed, and a two-stream instability was identified. 
Landau damping at $T = 0$ was also considered. We have shown that a residual temperature limit $T_r$ exists, associated with the  finite size of BECs. Finally, atom trapping and atom diffusion in velocity space were discussed, and possible extensions of the Landau damping theory were suggested. We believe that the present work illustrates the relevance  of the wave-kinetic description of dipolar BECs, and contributes  to the understanding of resonant atom-phonon interactions. This will eventually lead to a consistent model of quantum turbulence.

\begin{acknowledgements}

JTM and AG would like to thank the financial support of CNPq Brazil. AG also thanks funding of FAPESP Brazil. HT acknowledges FCT - Funda\c{c}\~{a}o da Ci\^{e}ncia e Tecnologia (Portugal) through the grant number IF/00433/2015.

\end{acknowledgements}


\begin{thebibliography}{99}

\bibitem{baranov}
M.A. Baranov, {\sl Phys. Rep.}, {\bf 464}, 71 (2008).

\bibitem{koch}
T. Koch, T. Lahaye, J. Metz, B. Fr\"ohlich, A. Griesmaier, T. Pfau, {\sl Nature Phys.}, {\bf 4}, 218 (2008).

\bibitem{pethick}
C.J. Pethick and H. Smith, {\sl Bose-Einstein Condensates in Dilute Gases}, 2nd ed., Cambridge University Press (2008).

\bibitem{odell}
D. H. J. O' Dell, S. Giovanazzi, and G. Kurizki, {\sl Phys. Rev. Lett.}, {\bf 90}, 110402 (2003).

\bibitem{santos}
L. Santos, G. V. Shlyapnikov, and M. Lewenstein, {\sl Phys. Rev. Lett.}, {\bf 90}, 250403 (2003).



\bibitem{fu}
H. Fu, Y. Wand and B. Gao, {\sl Phys. Rev. A}, {\bf 67}, 053612 (2003).

\bibitem{sabari}
S. Sabari, K. Prosezian and P. Muruganandam, {\sl Chaos, Solitons \& Fractals}, {\bf 103}, 232 (2017).

\bibitem{lee}
T.D. Lee and C.N. Yang, {\sl Phys. Rev.}, {\bf 105}, 1119 (1957); \,
T.D. Lee, K.W. Huang and C.N. Yang, {\sl Phys. Rev.}, {\bf 106}, 1135 (1957).

\bibitem{lima}A. R. P. Lima and A. Pelster, {\sl Phys. Rev. A}, {\bf 84}, 041604(R) (2011).
% Quantum fluctuations in dipolar Bose gases

\bibitem{pfau} 
M. Schmitt, M. Wenzel, B. B\"ottcher, I. Ferrier-Barbut, T. Pfau,
{\sl Nature}, {\bf 539}, 259 (2016),
%Self-bound droplets of a dilute magnetic quantum liquid

\bibitem{fabrocini}
A. Fabrocini and A. Polls, {\sl Phys. Rev. A}, {\bf 60}, 2319 (1999).

\bibitem{chomaz} L. Chomaz {\it et al.}, {Observation of the Roton Mode in a Dipolar Quantum Gas}, arXiv:1705.06914

\bibitem{stringari}
L.P. Pitaevskii and S. Stringari, {\sl Phys. Lett. A}, {\bf 235}, 398 (1997).

\bibitem{mend2005}
J.T. Mendon\c ca, P.K. Shukla, R. Bingham, {\sl Phys. Lett. A}, {\bf 340} 355 (2005).

\bibitem{natu}
S.S. Natu and R.M. Wilson, {\sl Phys. Rev. A}, {\bf 88}, 063638 (2013).

\bibitem{gardiner}
S.A. Gardiner, {\sl Phys. Rev. A}, {\bf  62}, 023612 (2000).

\bibitem{mend2016}
J.T. Mendon\c ca, {\sl J. Phys. A: Math Theor.}, {\bf 49}, 275501 (2016).

\bibitem{book}
J.T. Mendon\c ca and H. Ter\c  cas, {\sl Physics of Ultra-Cold Matter}, Springer, New York (2011).

\bibitem{fisher}
U.R. Fisher, {\sl Phys. Rev. A}, {\bf 73}, 031602 (2006).

\bibitem{tercas}
H. Ter\c cas, J.T. Mendon\c ca and G.R.M. Robb, {\sl Phys. Rev. A}, {\bf 79}, 065601 (2009).

\bibitem{giovanazzi}
S. Giovanazzi, A. G\"orlitz and T. Pfau, {\sl Phys. Rev. Lett.}, {\bf 89}, 130401 (2002).

\bibitem{kumar}
R.K. Kumar, T. Sriraman, H. Fabrelli, P. Muruganandam and A. Gammal, {\sl J. Phys. B}, {\bf 49}, 155301 (2016).

\bibitem{brodin}
G. Brodin, J. Zamanian and J.T. Mendon\c ca, {\sl Phys. Scr.}, {\bf 90}, 06820 (2015).

\bibitem{daligault}
J. Daligault, {\sl Phys. Plasmas}, {\bf 21}, 040701 (2014).

\bibitem{dysprosium1}
M. Lu, N. Q. Burdick, S. H. Youn, and B. L. Lev, {\sl Phys. Rev. Lett.}, {\bf 107}, 190401 (2011).

\bibitem{dysprosium2}
M. Raghunandan, C. Mishra, K. Lakomy, P. Pedri, L. Santos, and R Nath, Phys. Rev. A {\bf 92}, 013637 (2015).

\bibitem{mend2010}
J.T. Mendon\c ca, {\sl Phys. Rev. A}, {\bf 81}, 023421 (2010).


\end{thebibliography}
\end{document}